\documentclass[letterpaper, conference]{IEEEtran}
\usepackage{amsmath,amsfonts}
\usepackage{algorithmic}
\usepackage{algorithm}
\usepackage{array}
\usepackage[caption=false,font=normalsize,labelfont=sf,textfont=sf]{subfig}
\usepackage{textcomp}
\usepackage{stfloats}
\usepackage{url}
\usepackage{verbatim}
\usepackage{graphicx}
\usepackage{cite}
\usepackage{xcolor}
\usepackage{enumitem}
\makeatletter

\begin{document}

\title{Adversarial Security and Differential Privacy in mmWave Beam Prediction in 6G networks}

\makeatletter
\def\ps@IEEEtitlepagestyle{
  \def\@oddfoot{\mycopyrightnotice}
  \def\@evenfoot{}
}
\def\mycopyrightnotice{
  {\footnotesize
  \begin{minipage}{\textwidth}
  \centering
  Copyright~\copyright~2023 IEEE. Personal use of this material is permitted. Permission from IEEE must be obtained for all other uses, in any current or future media including reprinting/republishing this material for advertising or promotional purposes, creating new collective works, for resale or redistribution to servers or lists, or reuse of any copyrighted component of this work in other works.
  \end{minipage}
  }
}


\IEEEpubid{0000--0000/00\$00.00~\copyright~2021 IEEE}

\author{\IEEEauthorblockN{
\large Ghanta Sai Krishna\IEEEauthorrefmark{1}, Kundrapu Supriya \IEEEauthorrefmark{1}, Sanskar Singh\IEEEauthorrefmark{1}, Sabur Baidya\IEEEauthorrefmark{2}\\}
\IEEEauthorblockA{\IEEEauthorrefmark{1}\normalsize Department of Computer Science, IIIT Naya Raipur, India}
\IEEEauthorblockA{\IEEEauthorrefmark{2}\normalsize Department of Computer Science and Engineering, University of Louisville, KY, USA}
\normalsize {e-mail:  ghanta20102@iiitnr.edu.in, kundrapu20100@iiitnr.edu.in, sanskar21102@iiitnr.edu.in, sabur.baidya@louisville.edu}
}

\maketitle

\vspace{-10mm}
\begin{abstract}
In the forthcoming era of 6G, the mmWave communication is envisioned to be used in dense user scenarios with high bandwidth requirements, that necessitate efficient and accurate beam prediction.
Machine learning (ML) based approaches are ushering as a critical solution for achieving such efficient beam prediction for 6G mmWave communications. 
However, most contemporary ML classifiers are quite susceptible to adversarial inputs. Attackers can easily perturb the methodology through noise addition in the model itself. To mitigate this, the current work presents a defensive mechanism for attenuating the adversarial attacks against projected ML-based models for mmWave beam anticipation by incorporating adversarial training. 
Furthermore, as training 6G mmWave beam prediction model necessitates the use of large and comprehensive datasets that could include sensitive information regarding the user's location,  
differential privacy (DP) has been introduced as a technique to preserve the confidentiality of the information by purposefully adding a low sensitivity controlled noise in the datasets. It ensures that even if the information about a user location could be retrieved, the attacker would have no means to determine whether the information is significant or meaningless. With ray-tracing simulations for various outdoor and indoor scenarios, we illustrate the advantage of our proposed novel framework in terms of beam prediction accuracy and effective achievable rate  while ensuring the security and privacy in communications.

\end{abstract}


\section{Introduction}


The advancement in the modern emerging applications with the need of ultra high bandwidth and low latency has propelled the rapid progress in the cellular wireless communication technologies in contemporary years. Millimeter-wave (mmWave) came up as one of the major technologies in the 5G cellular communications supporting this high quality-of-service (QoS) demands, using massive Multiple-Input Multiple-Output (massive-MIMO) employing dense smaller antenna clusters. Now, as the cellular technologies are progressing towards 6G, incorporating artificial intelligence (AI)~\cite{9237460} supporting predictive communications and resource allocations become apparent including in mmWave beamforming.
The classic beam search is more complicated since each antenna element must be controlled. To reduce the complexity of beam search, a beam codebook can be used with predictive information. One of the main challenges in the mmWave beamforming communications is improving bandwidth efficiency while taking into consideration the issue of reducing cost and signal loss \cite{8947954} and finding the optimal mmWave beam \cite{8023460}. 
Several studies have been conducted recently in the area of mmWave beamforming communications, depending on the need and limitations.

\begin{figure}[!t]
\vspace{-2mm}
\centerline{\includegraphics[height=42mm,width=78mm]
{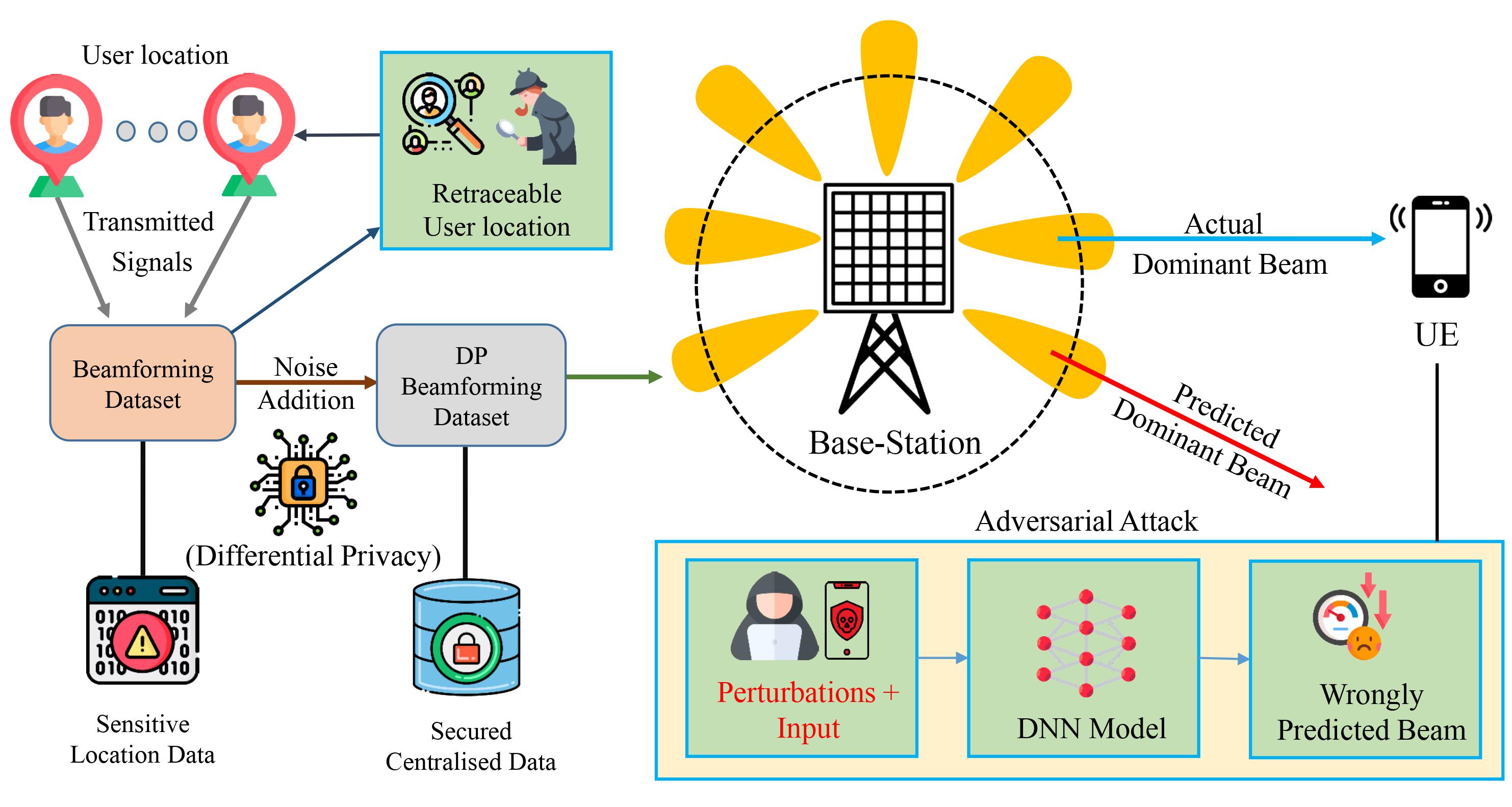}}
\vspace{-2mm}
\caption{Impacts of Adversarial Attacks and Role of Differential Privacy}
\label{fig1:framework}
\vspace{-6mm}
\end{figure}

Beamforming prediction using deep learning (DL) 
is susceptible to attacks using adversarial machine learning \cite{9609969}. The architecture of 5G networks is fraught with difficulties \cite{6812298}, one of which is a security system for beamforming prediction. Furthermore, different beamforming models in next generation (i.e. 6G)  are difficult to design and implement. 
In order to increase the overall system efficiency in 6G solutions, as AI-based methods, particularly deep learning (DL) techniques \cite{8395149} are used,
security concerns arise due to adversarial attack.
The current studies in this domain focus primarily on developing AI models, 
but security issues have  still been a major challenge. 
Privacy of the user's location is another  major concern while the transmission of the beams. Differential Privacy \cite{10.1145/2976749.2978318} can be employed during training of the ML model so that anonymity of the user can be maintained. 
Herein, we create a novel ML-based framework (fig.~\ref{fig1:framework}) to jointly train for adversarial security and differential privacy for mmWave beam prediction and demonstrate the performance of our framework in various communication scenarios. 

The main contributions of this work are summarized below.
\begin{itemize}
    \item To the best of our knowledge this is the first work on adversarial security and differential privacy together on the mmWave beam prediction.

    \vspace{1mm}
    \item We identify the tradeoff between the degree of adversarial security and differential privacy by introducing controlled noise in joint training with a novel ML-based framework.

    \vspace{1mm}
    \item We evaluate our framework extensively in various indoor and outdoor senarios through ray-tracing simulations. 
\end{itemize}


\section{Related Works}
A wide range of attacks have been implemented in recent years to study adversarial ML in telecommunications which includes beamforming\cite{9869669}, beam selection \cite{9471834} and channel estimation \cite{9745826}. 
In the past, 
adversarial attacks have been performed and analyzed using a variety of image datasets, but more recently, the Deep-MIMO dataset~\cite{deepmimo}, which was primarily created for mmWave and large MIMO systems, has been utilized to generate adversarial attacks. 

The majority of the research in the pertinent literature has focused on mmWave communication, developing various beamforming techniques and performance while ignoring issues with security and privacy. Yang et al. \cite{Yang20196GWC} presented a sophisticated 6G network design with AI assistance that can handle a number of services including discovery, automatic network adjustment, smart service provisioning, etc. Additionally, it covers AI-based techniques for effectively enhancing network efficiency in 6G networks, such as  mobility, and spectrum management. 
Catak et al.
\cite{article} presented an adversarial training-based mitigation technique for adversarial attacks against 6G ML models for mmWave beam prediction. The transmitted signal has been adjusted with a modified Fast Gradient Signed Method (FGSM) attack. The findings of adversarial training on two distinct models — defended and undefended, show that the defended model under attack has nearly identical mean square errors (MSE) to the undefended model not under attack.
There hasn't been any research on differential privacy in this area up to this point, but it has been extensively studied in the wireless communications industry. The authors in \cite{tian2022privacy}, proposed a fresh split learning (SL) approach that doesn't upload the raw data throughout inference and learning and solves the beam selection problem. To improve generalisation reliability and efficiency to quasi identically distributed data, it leverages the based feature mix approach. The evaluation's findings shown that, the suggested MPSL outperforms the conventional SL and federated learning approach in terms of test performance.

In contrast to the above  studies, our work considers stronger whitebox attack techniques than FGSM such as MI-FGSM, PGD and Iterative PGD attacks while showcasing the impacts of attacks in the 6G based beamforming. To this, defensive mechanisms are also introduced in order to mitigate the influence of adversarial perturbations on classifier performance. 
The previously proposed solutions also tend to rely heavily on the prior knowledge of user's location, which allows attackers to deduce sensitive information from the dataset. Herein, Differential Privacy (DP) is applied to mathematically induce a noise in the anticipated beam so that resulting beam is still accurate enough to generate aggregate insights of user's location while also maintaining the privacy of individual users.

\section{System Model}
This section presents a deep learning-based coordinated beamforming mmWave system and  describes the system model's functionalities, including downlink transmission, effective achievable rate, and deep learning model. The key necessities of utilising the DL model are also highlighted.

\begin{figure}[!t]
\centerline{\includegraphics[height=35mm,width=80mm]
{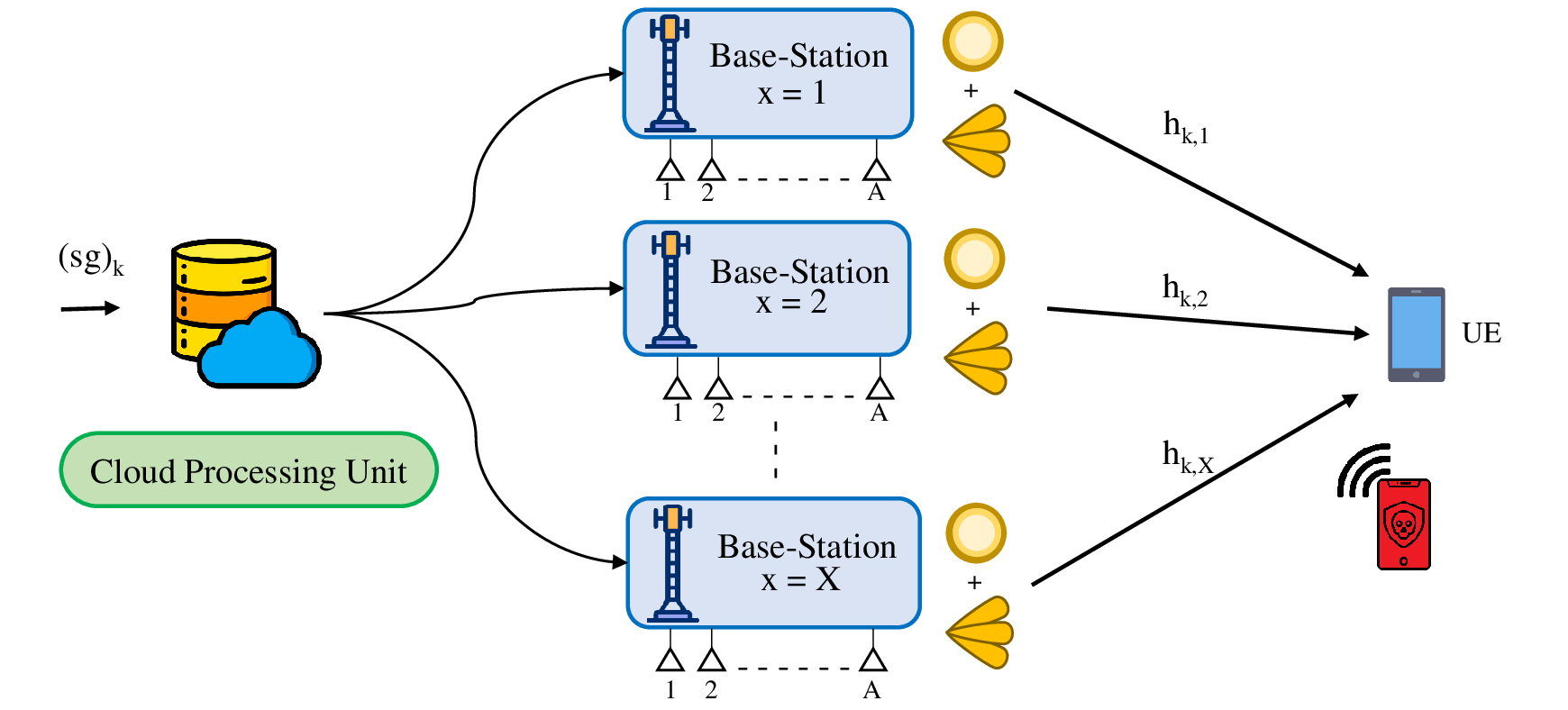}}
\vspace{-4mm}
\caption{Layout of the mmWave beamforming network}
\label{fig2:beamforming}
\vspace{-6mm}
\end{figure}

\subsection{Downlink Transmission}
Contemplate the mmWave communication system depicted in fig.~\ref{fig2:beamforming}, where X represents the count of base stations (BSs) with A antennas feeding user equipment(UE) or a mobile station with one antenna. These BSs are all linked to a cloud processing unit. A signal \textbf{sg} = $[sg_{1}, sg_{2}, . . . , sg_{K}]$ is transmitting with K additional secondary modulated signals (sub-carriers). In order to sustain transmission in multi-antenna wireless communications, a code vector $v_{k} = [v_{k,1}, v_{k,2}, . . . , v_{k,X}]^T$ is utilised to precode the transmitted signal with a digital precoder at the cloud processing unit. Furthermore, the K-point Inverse Fast Fourier Transform (IFFT) technique is employed to transform the retrieved transmitted signal into the time domain. The time-domain analogue beamforming $f_{x}$ (i.e., Beam steering vector) is applied by every BS and then transmits the resulting signal. Accordingly, the baseband signal at the $k$-th subcarrier from the $x$-th BS may be depicted as:

\vspace{-3mm}
\begin{equation}
\small
    \mathbf{b}_{k, x}={f}_{x} v_{k, x} (sg)_{k}
\end{equation}
\begingroup
\setlength{\tabcolsep}{5pt}

The downlink retreived signal at the kth subcarrier is denoted by:
\vspace{-3mm}
\begin{equation}
\small    \mathbf{g}_k=\sum_{x=1}^X \mathbf{h}_{k, x}^T \mathbf{b}_{k, x}+j_k
\vspace{-1mm}
\end{equation}

where $j_k$ denote the additive white Gaussian noise having variance as $\sigma^2$ for $k$th subcarrier. The channel vector between $x$th BS and the user is depicted through $\mathbf{h}_{k, x}$.

\subsection{Deep Learning to estimate RF beamforming vectors}
\vspace{-1mm}
Deep learning (DL) has lately been employed in physical layer communications applications \cite{8663966} such as signal identification, channel estimation and  precoding design and CSI feedback.
Utilizing the potential advantages of DL techniques in wireless communications results in a sophisticated strategy for training of massive MIMO channels  and scanning of a huge proportion of narrow beams. The DL model anticipates the BSs beamforming vectors straight from the retrieved signal at the scattered BSs with the utilization of omni or quasi-omni beam patterns. These are utilized to examine the diffraction and reflection of the pilot signal.

\subsubsection{ Model Architecture} 
The proposed architecture consists of beamforming prediction models to be located within the BSs in our system model. The cloud is simply in charge of precoding and transmitting the sent signals to the BSs in order to devise a less complex solution. For each beam coherence time $T_{BC}$,sequences of recurrent uplink training pilot of the form 
$\left\{s_k^{\text {pilot }}\right\}_{k=1}^{K}$ is transmitted by the user. Beam coherence time is referred to as the average period that the beam remains aligned. The retrieved pilot sequences are combined on a radio frequency beamforming vector and sent to the cloud via BSs. Pilot signals are used to characterize the prediction channel. The cloud feeds the omni-retrieved signals $C_{k, x}^{\text {omni }}, \forall x$ from all of the BSs into the DL algorithm and instructs the pretrained neural network to determine the optimal beamforming vector which further tries to boost the potential rate as in  
\vspace{-2mm}
\begin{equation}
\small
    R_x^{(p)}=\frac{1}{K} \sum_{k=1}^K \log _2\left(1+\mathrm{SNR}\left|\mathbf{h}_{k, x}^T \mathbf{o}_p\right|^2\right)  
\vspace{-1mm}
\end{equation}

where, $h_{k,x}$ represents the channel coefficient matrix for xth BS at the kth subcarrier and $o_p$ illustrate the channel coefficient for omni-beams. The DL model is trained in real-time to comprehend the inherent relationship among the rates of the various RF beamforming vectors and the 
Orthogonal Frequency-Division Multiplexing(OFDM) omni-retrieved signals collected simultaneously at each BS,  that further serve as a distinctive confirmation of the user's location.
Eventually, the BS terminals utilize the anticipated RF beamforming vectors $f_x^{\mathrm{DL}}$ to aggregate the uplink pilot sequences and predict the potent channel $\mathbf{h}_{k, x}^T {f}_x^{\mathrm{DL}},\forall n$, that are then employed to generate the beamforming vectors from the cloud baseband.

\subsubsection{DNN Training}
The Deep Neural Network employed in the present work comprises of 7 fully-connected feed forward layer with ReLU as the activation function. It is worth noting that the output of each dense layer is normalized batch-wise before being sent to the subsequent layer. A layer is introduced with drop-out to guarantee regularisation and prevent the chance of over-fitting in the neural network. The DNN architecture implemented in the current study is shown in the fig.~\ref{fig:DL_framework}.

\begin{figure}[!t]
\centerline{\includegraphics[height=40mm,width=80mm]{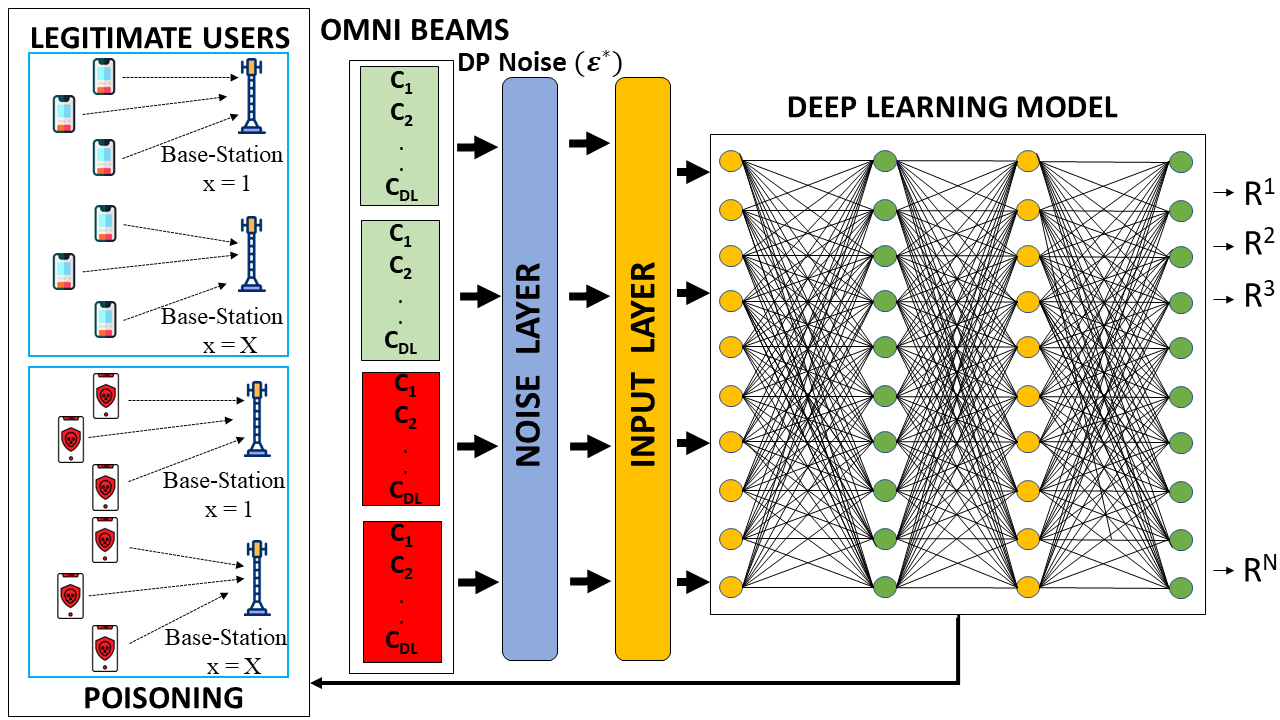}}
\vspace{-3mm}
\caption{Redeemed omni beams to the DL based codeword prediction model}
\label{fig:DL_framework}
\vspace{-4mm}
\end{figure}

\vspace{-2mm}
\subsection{Effective Achievable Rate}
 Ideal channel estimation knowledge fulfills the best attainable rate; yet, because due to the huge amount of antennas, channel state information necessitates a significant training overhead. Also, as the user traverses, the RF beamforming vector and channel information must be revised. The channel beam coherence time, $T_B$ and channel coherence time $T_C$ could be utilized to address these concerns. Furthermore, the beams remain aligned during the length of beam coherence $T_B$ and $T_B$ is often larger than $T_C$. For users with greater mobility, the time period $T_B$ reduces, resulting in a reduced data throughput for the same beam training overhead and beamforming codebook vectors. The first $T_t$ is required for beamforming drafting and training the channel, while the remainder is dedicated to data transmission. Hence, the effective achievable rate can be elucidated as follows.

\vspace{-4mm}
\begin{equation}
\small
    R_{e f f}=\left(1-\frac{T_{t}}{T_B}\right) \sum_{k=1}^K \log _2\left(1+S N R\left|\sum_{x=1}^X \mathbf{h}_{k, x}^T {f}_{x} v_{a, x}\right|^2\right)
\vspace{-1mm}
\end{equation}

The effective achievable rate (EAR) must be maximized to construct a model having coherent channel training and beamforming architecture. Therefore, the final problem formulation can be defined as :

\vspace{-2mm}
\begin{equation}
\small
\prod\left(T_t,\left\{\mathbf{v}_k\right\}_{k-1}^K, \mathbf{F}^R, \mathcal{C}\right)=\operatorname{argmax}\left(R_{e f f}\right)
\end{equation}

\vspace{-6mm}
\begin{equation}
\small
    \begin{gathered}
\text { s.t. } {f}_x \in \mathcal{C}, \forall x
\end{gathered}
\end{equation}

\vspace{-6mm}
\begin{equation}
\small
    \left\|v_k\right\|^2=1 \quad \forall k
    \vspace{-1mm}
\end{equation}

Here $\mathcal{C}$ represents the quantized codebook for the BSs RF beamforming vectors and $\mathbf{F}^R$ is the RF precoding matrix, $\mathbf{F}^R F=\operatorname{blkdiag}\left(\mathbf{v}_1, \mathbf{v}_{2}, \ldots, \mathbf{v}_X\right) \in \mathbf{V}^{X A \times X}$.  Working through these equations culminates in an answer for a minimal channel training overhead and the generation of the beamforming vectors to yield the highest attainable rate.

\section{Research Methodology}

\vspace{-1mm}
\subsection{Adversarial Attack Models}
Adversarial ML is an attack strategy to deceive neural network(NN) models by providing input that has been cleverly altered with a slight change
to generate a breakdown inside the models. The attacker subtly modifies the query to obtain the desired result from a production-deployed model instead of changing the training instances. This is a violation of model input integrity, which results in fuzzing-style attacks, compromising the model's classification performance.

Recently, numerous white-box adversarial attacks have been proposed to alter models’ prediction by introducing minute
carefully designed perturbations. Some of the state-of-the-art (SOTA) attack practice (i.e. FGSM \cite{https://doi.org/10.48550/arxiv.1412.6572}, Momentum Iterative FGSM, PGD \cite{https://doi.org/10.48550/arxiv.1706.06083} and Iterative PGD) will be utilized to evaluate the performance of defense models in this paper, which will be briefly introduced as follows.

\subsubsection{Fast Gradient Sign Method (FGSM) attack} It is a typical one-step attack algorithm designed to target neural networks by utilizing the training gradients. 
Here, instead of reducing the loss by altering the weights based on the backpropagated gradients, 
it adjusts the input data $y$ to maximize the adversarial loss $L(\theta, y, z)$ based on the same backpropagated gradients for some model parameter $\theta$. The FGSM-attack 
can be formulated as follows:
\vspace{-1mm}
\begin{equation}
\small
    y_{t+1}=y_t+\epsilon \cdot \operatorname{sign}\left(\nabla_{y_t} L\left(\theta,y_t, z\right)\right),
\end{equation}
 
where z is the ground truth label for $y$, $y_{t+1}$ is the new adversarial instance that maximizes the classifier hypothesis loss and $\epsilon$ denotes the magnitude of the perturbation applied. 

\vspace{1mm}
\subsubsection{Momentum Iterative FGSM attack} 
Iterative FGSM \cite{https://doi.org/10.48550/arxiv.1607.02533} is an improvised version of FGSM which apply gradient updates iteratively.
MI-FGSM is more transferable than I-FGSM since it accumulates the gradients of each iteration to stabilize the update direction and avoid superfluous local maxima.
\begin{equation}
\small
    \begin{gathered}
    \vspace{-1mm}
G_t=\mu \cdot G_{t-1}+\frac{\nabla_{y_t} L\left(\theta,y_t, z\right)}{\left\|\nabla_{y_t} L\left(\theta,y_t, z\right)\right\|}, \\
y_{t+1}=y_t+\alpha \cdot \operatorname{sign}\left(G_t\right)
\end{gathered}
\end{equation}
Here, $G_{t-1}$ is the cummulative gradient up to the $(t-1)$-th iteration with $G_0=0$ and a decay factor of $\mu$ . Also, $\alpha = \epsilon/T$ represents a small step size, and T is the number of iterations.

\vspace{1mm}
\subsubsection{Projected Gradient Descent attack} PGD could well be regarded of as an enhanced variant of I-FGSM being devoid of constraint $\alpha$. The PGD projects the adversarial samples learned from each iteration onto the benign samples' neighbour in order to restrict the adversarial perturbations. As a result, the adversarial perturbation size is less than $\epsilon$.
\begin{equation}
 \small   y_{t+1}=\operatorname{Proj}\left\{y_t+\alpha \cdot \operatorname{sign}\left[\nabla_y L\left(\theta, y_t, z\right)\right]\right\}
\end{equation}
where $Proj$ gives the projection of the revised adversarial sample onto the neighbour and an acceptable range.
Iterative PGD (IPGD) is an extended version of PGD that seeks to apply PGD repeatedly during the network's training phase.

\subsection{Adversarial Training}
Adversarial training (AT) is an inherent defensive approach against adversarial samples to increase the resilience of a neural network by training it with adversarial examples, or on mix of clean and adversarial samples. AT strategy is shown to be coherent while defending the models from adversarial attacks \cite{https://doi.org/10.48550/arxiv.1706.06083}. It can be well generalised as a min-max optimization problem which can be elucidated as:
\vspace{-1mm}
\begin{equation}
\small
    \min _\theta \underset{(y_t, z)}{\mathbb{E}}\left(\max _{D\left(y_t, y_{t+1}\right)<\epsilon} L\left(h_\theta\left(y_{t+1}\right),h(y_t\right)\right)
\vspace{-1mm}    
\end{equation}

\vspace{-1mm}
\begin{equation}
\small
    \begin{gathered}
\text { s.t. } h_\theta\left(y_{t+1}\right) \neq z \quad \wedge D\left(y_t, y_{t+1}\right)<\epsilon
\end{gathered}
\end{equation}

where $D()$ represents distance metric between initial input $y_t$ and it's adversarial variant $y_{t+1}$ while $h_\theta()$ depicts the neural network based classifier model.
Thus, once the model is trained under several attacks, adversarial samples are generated by the model, that are merged with information from genuine 'good' users and fed again to the training cycle. The training procedure is finished when the model achieves the stable state.



\subsection{Differential Privacy}
DP is a mathematical strategic tool for purposely introducing noise into a dataset, ensuring plausible deniability to any individual whose data may be exploited to harm them while yet being able to compute desired statistics with high accuracy.

Consider any two beamforming databases as $d$ and $d^{\prime}$ taken from DeepMIMOv2, that differ precisely in one record of user location. If $K_f$ is a function used to answer to location query $f$, then $K_f$ provides $\epsilon^{\ast}$-differential privacy if with probability $1$, for any $r \subseteq \operatorname{Range}\left(K_f\right)$ the following occurs:

\vspace{-2mm}
\begin{equation} 
\small
\operatorname{Pr}\left[K_f(d) \in r\right] \leq \mathrm{e}^{\varepsilon^{\ast}} \operatorname{Pr}\left[K_f\left(d^{\prime}\right) \in r\right]
\end{equation}

Here, $\epsilon^{\ast}$ is the privacy budget which specifies an upper-bound on privacy loss of user's location information. 
This  work employs DP-SGD which achieves privacy by clipping the gradients and then adding the noise proportionally.

\section{Experimental Results and Analysis}

\subsection{Radio Frequency Beamforming Data Generation}
We evaluate our proposed framework using the DeepMIMOv2 dataset
which is explicitly determined by the ray tracing scenarios and the parameters set. A ray-tracing scenario ‘R’ incorporates a range of access points/BSs and clients that are spatially distributed inside an indoor or outdoor surroundings. BSs and clients in the ray-tracing scenario often contain omni or quasi-omni transmitters. The ray-tracing simulation produces performance indicators for each stream between every sender and receiver antennas. An enormous amount of base stations and clients have been included in the ray-tracing scenarios so that we can increase the size of the dataset for machine learning applications. 
To generate the dataset, we must provide two inputs to the data generation code: ray-tracing scenarios and the parameter set as mentioned already. 
\subsubsection{Ray-Tracing Scenarios}We extract the ray-tracing outcomes for the DeepMIMOv2 data using Wireless InSite by Remcom's precise ray-tracing simulator. Ray-tracing simulations produce channel parameters which accurately reflect the placements of transmitter, components, geometry of the environment, etc. 
\subsubsection{Dataset Parameters} Various communication parameters, e.g., number of active base stations, active users, antenna spacing, bandwidth, etc are variable in ray-tracing scenarios. 
The {\bf Table~\ref{table:comm_params}} represents the parameters that have been used for generating dataset from different ray-tracing scenarios.

\begin{table*}[!t]
\scriptsize
	\centering
	\caption{Parameters used in different scenarios in DeppMimoV2 data}
 \vspace{-2mm}
	\label{table8}
	\begin{tabular}{|c|p{32mm}|c|c|c|c|c|c|c|c|c|}
	\hline
Parameters & Description & O1 Scenario & O1 Blockage Scenario & O2 Scenario & I1 Scenario & I2 Scenario & I3 Scenario\\ 
\hline
\centering Active BS & Base stations to be activated & 1,2 & 1,2 & 1,2 & 1 & 1 & 1,2\\
\hline
 Active Users & Activating a group of users & 100 -300 & 100-700 & 700 - 800 & 200 - 400 & 100 - 500 & 600 - 1000\\
\hline
\centering Number of BS Antennas & Number of antennas & 1,32,8 & 1,32,8 & 1,32,8 & 1,32,8 & 1,32,8 & 1,32,8\\
\hline
Antenna Spacing & Distance between the base station transimitter gird elements & 0.5 & 0.5 & 0.5 & 0.5 & 0.5 & 0.5\\
\hline
 Bandwidth & Bandwidth in Gigahertz & 0.5 & 0.5 & 0.5 & 0.5 & 0.5 & 0.5\\
\hline
Number of Paths &  5 paths w/ highest recv. power & 5 & 5 & 5 & 5 & 5 & 5\\
\hline
\end{tabular}
\label{table:comm_params}
\vspace{-4mm}
\end{table*}

\begin{figure}[!t]
\centerline{\includegraphics[width=0.80\linewidth]{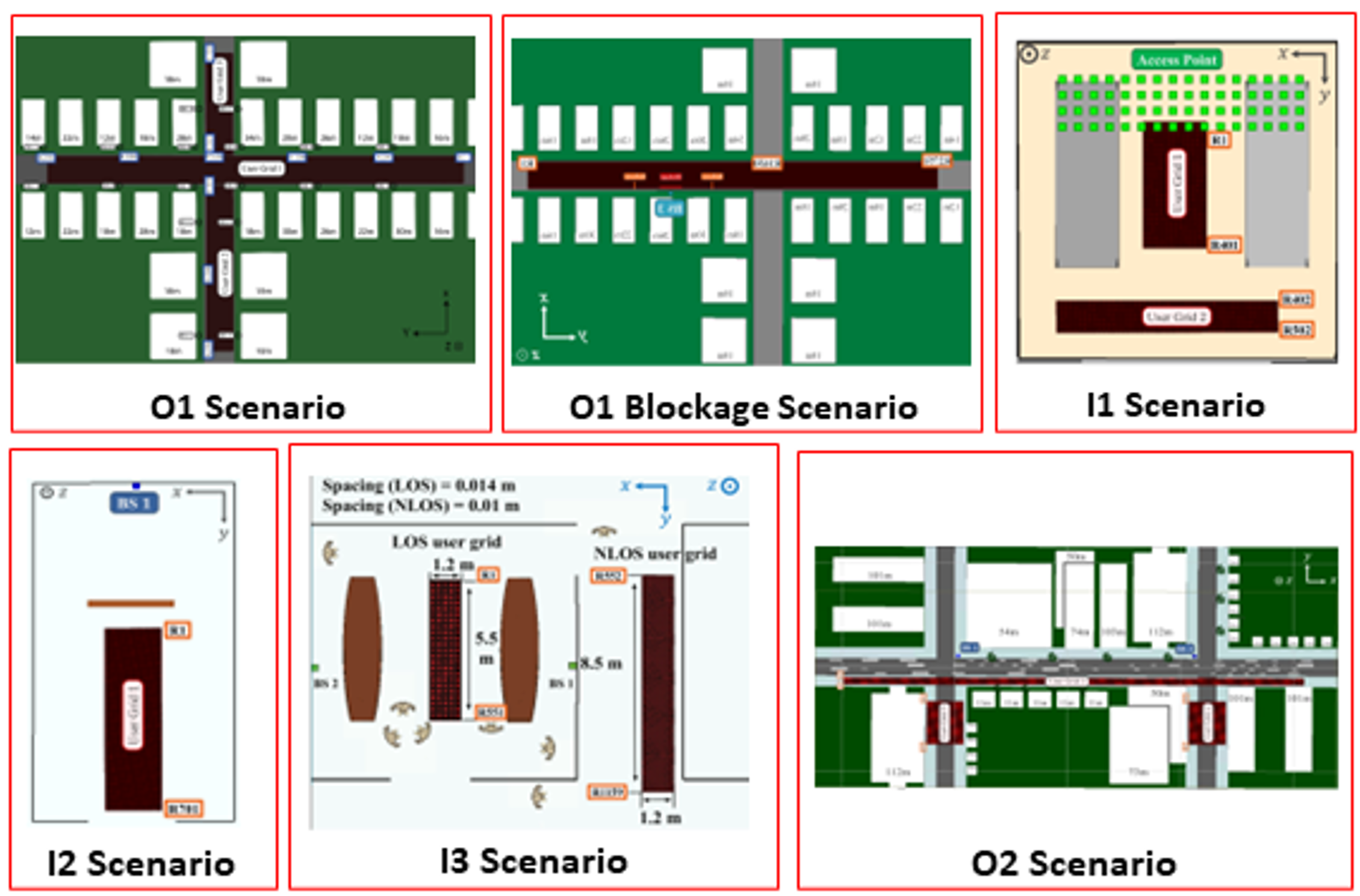}}
\vspace{-2mm}
\caption{Top-View of all the Scenarios~\cite{deepmimo}}
\label{fig:scenarios}
\vspace{-6mm}
\end{figure}

\subsection{Experiments on Different Cases}
The proposed model has been evaluated on the basis of 2 separate cases for all the 6 individual scenarios shown in fig.~\ref{fig:scenarios}. The varied cases can be illustrated as:
\begin{itemize}[leftmargin=4mm]
  \item \textbf{Non-Private Model under attacks (C1) :}  This case reflects  various attack methodologies applied on the undefended non-private beamforming codebook prediction model. Then the  undefended model is trained adversarially to explore the variances obtained in training phase.
  \item \textbf{ Private Model under attacks (C2):} This case examines the same attack techniques employed over an defended DP model. Similar to C1, it also draws the impact of the attacks after the undefended DP model has been defended using adversarial training.
\end{itemize}

Hence, two models (i.e. non-private and private) were developed corresponding to the above mentioned respective case instances. The optimum set of tuned hyper-parameters are employed in both the model's architecture are displayed in {\bf Table~\ref{table:hyperparams}}.

\begin{table}[h]
\vspace{-4mm}
\caption{Tuned Hyper-parameters for the  beam prediction model}
\label{table2}
\vspace{-4mm}
\begin{center}
\scalebox{0.8}
{
\begin{tabular}{|c|c|c|}
\hline
\textbf{Hyper-parameter} & \textbf{Without Privacy} & \textbf{With Privacy}  \\ 
\hline
Number of beams & 512 & 512\\
\hline
Epochs & 10 & 15\\
\hline
Batch Size & 100 & 100\\
\hline
Dropout Rate & 0.05 & 0.05\\
\hline
Learning Rate & 0.25 & 0.0005\\
\hline
Hidden Layers & 4 & 4\\
\hline
Optimizer & RMSprop & DP-SGD\\
\hline
\end{tabular}
}
\label{table:hyperparams}
\vspace{-4mm}
\end{center}
\end{table}

\begin{figure*}[ht]
\centerline{\includegraphics
[width=0.99\linewidth]
{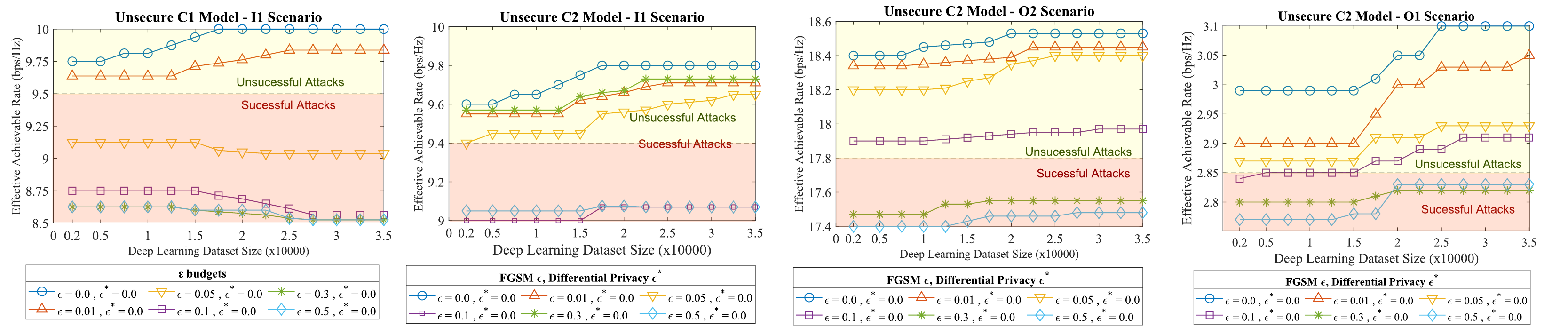}}
\vspace{-2mm}
\caption{Effective Achievable Rate of RF beamforming DL model in all the Scenarios}
\label{EAR}
\vspace{-2mm}
\end{figure*}

\begin{figure*}[ht]
\centerline{\includegraphics[width=0.99\linewidth]{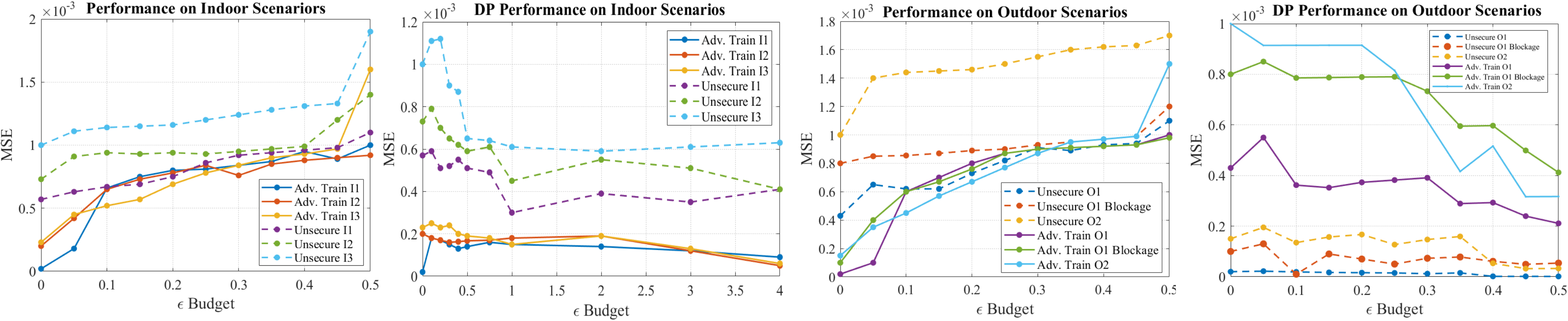}}
\vspace{-2mm}
\caption{MSE in RF beamforming DL model in all the Scenarios}
\vspace{-4mm}
\label{MSE in RF model}
\label{fig:scenarios_resuts}
\end{figure*}

Approximately thirty-five thousand non-perturbed instances are utilized for training both models. The experimental analysis of adversarial attacks and differential privacy has been performed regarding Effective Achievable Rate (EAR) and Mean Squared Error (MSE). The EAR has been recorded for private and non-private models under 6 FGSM attacks with variable parameters in various scenarios and variable training dataset size, as illustrated in Fig.~\ref{EAR}. In the unsecure C1, four  FGSM attacks were successful, whereas, in secure C2, only two were successful in the I1 scenario. Similar results have been observed in the other scenarios, which depict that the private model has more capability to resist adversarial attacks. 

Fig.~\ref{MSE in RF model} illustrates the obtained MSE scores under $\epsilon$ budget in indoor and outdoor scenarios with FGSM attack instances for both cases (C1 and C2). The plot shows that after a specific $\epsilon$ value, the adversarially trained RF beamforming model's MSE values become quasi-static. On the contrary, the MSE values of the undefended models remain elevated over the MSE values of adversarially trained models irrespective of the scenario. In differentially private models, unsecure and adversarially trained performance is distinguishable compared to non-private models.

\subsection{Tradeoff of $\epsilon$ between AT and DP }
The present section discusses about the differences between both variant of noises described in this study. Introduction of noise ($\epsilon$) by the attacker implies a carefully computed petrurbations whose main task is to mislead the network. In this case, as the $\epsilon$ value gets larger, it becomes relatively easy to deceive the network. Nevertheless, as a result of this trade-off, the perturbations become more apparent.
$\epsilon^{\ast}$ notion has been mentioned in the context of DP where $\epsilon^{\ast}$ is a metric of privacy loss after a differential change is applied in beamforming dataset. 
A smaller $\epsilon^{\ast}$ will yield better privacy but less accurate response. The tradeoff relationship between the epsilon values for AT and DP have been showcased using C1 and C2 models MSE scores under $\epsilon$ budget in the fig.~\ref{Tradeoff}.
The model's performance will increase with an increase in the privacy budget, and will decrease with an increase in epsilon in indoor and outdoor scenarios. 

\begin{figure}[!t]
\centerline{\includegraphics
[width=1.05\linewidth]
{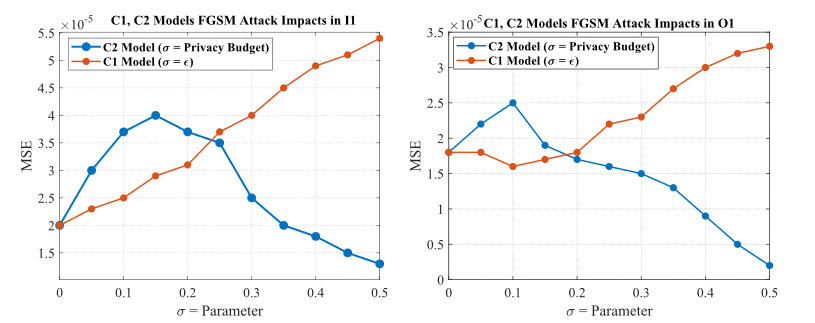}}
\vspace{-2mm}
\caption{Tradeoff of $\epsilon$ for C1 and C2 model}
\label{Tradeoff}
\vspace{-2mm}
\end{figure}

\begin{figure}[!t]
\centerline{\includegraphics
[width=0.90\linewidth]
{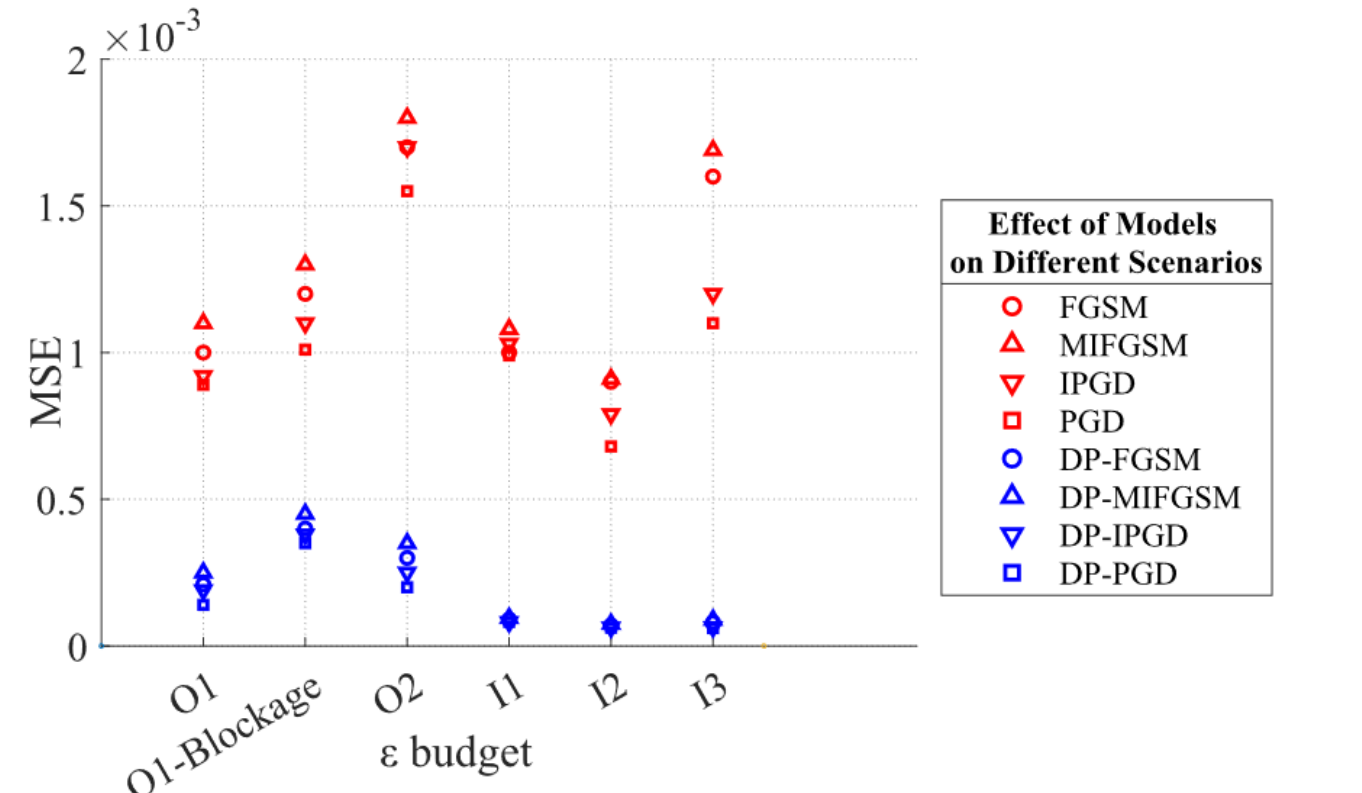}}
\vspace{-2mm}
\caption{Performance of Adversarially Trained Private and Non-Private Models }
\label{fig1}
\vspace{-6mm}
\end{figure}

Fig. \ref{fig1} represents the performance of Adversarially Trained Private and Non-Private Models on six-ray tracing scenarios. The differential privacy played a positive role in decreasing the impacts of adversarial attacks, the increase in performance with the differentially private model is approximately eight times the non-private model. This can be calculated by estimating the distance between the private and non-private data points in the scatterplot of Fig. \ref{fig1}.

\section{Conclusion}
The privacy and security of 6G mmWave beamforming are crucial due to its significant role in both present and future applications. Leveraging such architecture with DL subjects it to the danger of adversarial samples and user's location knowledge retrieval by attackers, which have not received proper attention in the pertinent literature. The present work uses DeepMIMOv2 dataset as a benchmarking tool in the domain of mmWave communication. Several indoor as well as outdoor scenarios of the ray-tracing simulations have been employed for the experiments. The study has been conducted by considering 2 separate cases to demonstrate the impact of DP in enhancing the confidentiality of user's location information. The procured results demonstrate that the initial model in both the cases are susceptible to various attack instances such as FGSM, MI-FGSM, PGD and Iterative PGD. To this, the beamforming DL model is iteratively supplemented by the adversarial samples. This mitigation strategy ensues that repetitive adversarial training proficiently improves the RF beamforming codebook anticipation performance and generates a much efficient predictive model.     

\section*{Acknowledgment}
This work is partially supported by the US National Science Foundation (EPSCoR \#1849213).

\bibliographystyle{IEEEtran}
\bibliography{references}

\vfill

\end{document}